\documentclass[aps,pra,twocolumn,superscriptaddress,longbibliography,reprint]{revtex4-1}
\setcitestyle{square}

\usepackage{graphicx}
\usepackage{enumerate}
\usepackage[colorlinks=true, linkcolor=blue, citecolor=blue]{hyperref}
\usepackage{textcomp}
\usepackage{amsmath}
\usepackage{amssymb}
\usepackage[english]{babel}
\usepackage[capitalise]{cleveref}

\newcommand{\etal}{\textit{et al.}}

\newcommand{\chain}{\mathrm{c} }
\newcommand{\Hc}{\mathrm{H.c.} }

\newcommand{\SIAM}{\scriptscriptstyle \mathrm{SIAM}}

\newcommand{\opa}[1]{{\hat{a}^{\phantom \dagger}}_{#1}}
\newcommand{\opadag}[1]{{\hat{a}^{\dagger}}_{#1}}

\newcommand{\opc}[1]{{\hat{c}^{\phantom \dagger}}_{#1}}
\newcommand{\opcdag}[1]{{\hat{c}^{\dagger}}_{#1}}

\newcommand{\opd}[1]{{\hat{d}^{\phantom \dagger}}_{#1}}
\newcommand{\opddag}[1]{{\hat{d}^{\dagger}}_{#1}}

\newcommand{\opn}[1]{{\hat{n}^{\phantom \dagger}}_{#1}}

\newcommand{\Ham}{\widehat{H}}
\newcommand{\Hloc}{\widehat{H}_{\mathrm{loc}}}
\newcommand{\Hhyb}{\widehat{H}_{\mathrm{hyb}}}
\newcommand{\Hcond}{\widehat{H}_{\mathrm{cond}}}

\newcommand{\step}[1]{\textcircled{\raisebox{-0.9pt}{#1}}}

\begin{document}

\title{Kondo nanomechanical dissipation in the driven Anderson impurity model}
\author{Lucas Kohn}
\affiliation{SISSA, Via Bonomea 265, I-34136 Trieste, Italy}
\author{Giuseppe E. Santoro}
\affiliation{SISSA, Via Bonomea 265, I-34136 Trieste, Italy}
\affiliation{International Centre for Theoretical Physics (ICTP), Strada Costiera 11, I-34151 Trieste, Italy}
\affiliation{CNR-IOM, Via Bonomea 265, I-34136 Trieste, Italy}

\author{Michele Fabrizio}
\affiliation{SISSA, Via Bonomea 265, I-34136 Trieste, Italy}

\author{Erio Tosatti}
\affiliation{SISSA, Via Bonomea 265, I-34136 Trieste, Italy}
\affiliation{International Centre for Theoretical Physics (ICTP), Strada Costiera 11, I-34151 Trieste, Italy}
\affiliation{CNR-IOM, Via Bonomea 265, I-34136 Trieste, Italy}

\begin{abstract}
The cyclic sudden switching of a magnetic impurity from Kondo to a non-Kondo state and back was recently shown to  involve an important dissipation of the order of several $k_BT_K$ per cycle. The possibility to reveal this and other electronic processes through nanomechanical dissipation  by e.g., ultrasensitive Atomic Force Microscope (AFM) tools currently represents an unusual and interesting form of spectroscopy. Here we explore the dependence on the switching time of the expected dissipation, a quantity whose magnitude is physically expected to drop from maximum to zero between sudden and slow switching, respectively. By applying a recently established matrix-product-state based time-dependent variational algorithm to the magnetic field-induced Kondo switching in an Anderson model of the magnetic impurity, we find that dissipation requires switching within the Kondo time scale $\hbar(k_B T_K)^{-1}$ or faster. While such a fast switching seems problematic for current AFM setups, the challenge is open for future means to detect this dissipation by time-dependent magnetic fields, electrostatic impurity level shift, or hybridization switching.
\end{abstract}

\date{13 May 2022}

\maketitle

\section{Introduction}
Interest in the Kondo effect, a paradigmatic single-site many-body phenomenon, is, 50 years after Anderson's theoretical breakthrough \cite{ PRL_Anderson_1969, PRB_Anderson_1970}, still unabated. Spectroscopically, zero-bias tunneling anomalies  \cite{PR_Appelbaum_1967} represent the well known spectroscopic signal of the Kondo effect in electron conductance through a spin-carrying site. 
Here we discuss fresher alternatives, provided by nanomechanics. 
A more recent concept is that atomic force microscope (AFM) mechanical dissipation can also serve as a spectroscopic tool, as for example has been shown with quantum dots\cite{PNAS_Cockins_2010,Nature_Kisiel_2018}. 
With increased sensitivity, the so-called pendulum AFM
tool enables extremely accurate yet non-invasive measurements of dissipation in oscillating tips experiments~\cite{PRB_Fuchs_1999, Nature_Kisiel_2011, Nature_Langer_2014, PRL_Meyer_2015}. At every cycle, a fraction of the vibrating tip's mechanical energy is dissipated through dynamical processes of all kinds, electronic and magnetic as well as ionic, going on in the sample underneath. The dissipated energy and its dependence upon parameters such as strength of interaction between tip and surface-adsorbed impurity, voltage, temperature, or magnetic field can provide spectroscopic evidence of processes such as electronic transition in quantum dots and surface unpaired electron centers  \cite{PNAS_Cockins_2010,Nature_Kisiel_2018}, besides collective phenomena such as normal-superconductor transitions \cite{Nature_Kisiel_2011}, or structural transitions\cite{PRL_Meyer_2015}.

More recently, some of us suggested that if the ``on-and-off'' switching of the Kondo effect could be operated by the tip itself in the course of each oscillation cycle, then a corresponding mechanical dissipation of order $k_B T_K$, where $T_K$ is the Kondo temperature, might be detectable in pendulum AFM measurements ~\cite{PRB_Baruselli_2017}. 
That mechanical cost must be provided to the vibrating tip so as to maintain its oscillation.  
In that first study, the Kondo switching was modeled by instantaneously turning on and off the hybridization interaction between the impurity and the free electrons in the metal. 
Such a sudden switch permits to express the dissipation in terms of equilibrium quantities~\cite{PRB_Baruselli_2017}.
A finite frequency modulation of hybridization was subsequently discussed in Ref.~\cite{ArXiv_Baruselli_2018}, confirming that periodic hybridization switching should contribute a Kondo dissipation per cycle by about $k_B T_K$, without much further ado.

Controlled on-off switching of the Kondo effect appears to have been realized, in combined STM/AFM experiments, probably by a shuttling proton chemically changing the spin of a surface-deposited molecular impurity between spin-1/2 (Kondo) and spin-1 (non-Kondo)~\cite{Science_Jacobson_2017}. 
In that case, the measured dissipation per cycle was much larger than $k_B T_K$. 
Yet, it was not clear what fraction, if not the whole, of the dissipation was due to the proton shuttling between tip and impurity -- a mechano-chemical effect, unrelated to Kondo. In principle, it could in fact be possible in the future
 to switch the impurity-metal hybridization e.g., by mechanically lifting the impurity off and on the metal substrate~\cite{ACS_Pawlak_2016}.\\
 
A less draconian possibility to disturb the Kondo effect of a surface impurity  by flying the tip over the impurity, could be just to shift the electronic impurity level, from e.g., well below Fermi to near or well above Fermi and back again, periodically. 
Yet another Kondo switching manoeuvre could be to expose the impurity to an oscillatory magnetic field -- for example, by a pendulum AFM tip carrying a permanent magnet. 
In such hypothetical experiments however, the on-off Kondo switching will not occur instantaneously -- a nonzero switching time must be required depending on parameters, including oscillation frequencies, amplitudes, interaction, etc. 
The Kondo dissipation effect described by Baruselli {\em et al.} ~\cite{PRB_Baruselli_2017} for infinitely fast switching needs therefore to be reconsidered and updated to account for finite switching times.\\

Here we calculate the nanomechanical dissipation caused by a gradual, time-dependent on-off switching of Kondo effect of a model impurity, therefore extending the infinitely sudden results of Ref.~\cite{PRB_Baruselli_2017}.
We focus on two separate cases, among the {\em gedanken} experiments listed above. 
Discarding the chemical mechanisms, we restrict to a time-dependent impurity level shift, and to a time-dependent magnetic field.  
For the first, we calculate the dissipation caused by a time-dependent evolution of the impurity energy level, thus modeling the effect of electrostatic tip-impurity interaction on the impurity electronic chemical potential during the oscillation. 
This is the mechanism at work in, e.g., Ref.~\cite{Nature_Kisiel_2018}. 
For the second, we will look for the dissipation effect caused by a time-dependent magnetic field such as could be exerted by the stray field of an oscillating ferromagnetic tip. 
That done, we need not consider the alternative case of a time-dependent switching of impurity-bath hybridization, because the evolution after Kondo is suppressed and restored, is expected in that instance to resemble closely the magnetic field case.

Anticipating the main outcomes of this study, we will first demonstrate how an oscillatory magnetic field perturbation -- much as will a large asymmetric impurity level oscillation, or as periodic on-off switching of the impurity hybridization -- should produce Kondo dissipation. The second decisive outcome concerns the switching time. 
No dissipation is expected to occur for an infinitely slow, adiabatic, Kondo switching, no matter how operated, while a sudden, infinitely fast, switching of hybridization --- such as that of Ref.~\cite{PRB_Baruselli_2017} --- will likely provoke the maximum possible dissipation, under reasonable assumptions of monotonicity. 
The question to answer is therefore double. 
First, we will study what will be the dissipation caused by oscillating the impurity energy level up and down (an electrostatic effect that could be exerted by an oscillating tip), or that caused by an oscillatory external magnetic field (an effect such as that caused by an oscillating ferromagnetic tip). 
Second, we should describe how would that dissipation drop when the switching velocity is gradually reduced from infinity, where it is likely largest, to realistically lower values.  
The answer to these questions will determine in which scenario it could be possible to observe signatures of the Kondo effect in experimentally accessible dissipation approaches.

We address the fundamental quantum mechanical starting point of these questions by simulating the full real-time dynamics of the externally perturbed Anderson model of a quantum impurity, whose Kondo effect is switched with different means and  time-dependencies. All results are obtained employing our recently established, state-of-the-art numerical technique based on matrix product states~\cite{PRB_Kohn_2021, arXiv_Kohn_2021} .
The remaining questions of experimental time scales, more qualitative, will be addressed in the discussion part.

The paper is organized as follows. 
Section \ref{sec:model} summarizes the single-impurity Anderson model, the time-dependent protocol, and the matrix-product-state techniques employed to study the full time-dependence of the model. Section \ref{sec:results} contains the results of our simulations where dissipation is calculated for the two mechanisms we have considered: a switch of the local impurity level, and a switch of a local magnetic field. 
Finally, Sec.~\ref{sec:conclusions} contains a final discussion and draws some conclusions. \\

\section{Model, protocol and method} \label{sec:model}
\subsection{Anderson model}

We shall calculate the dissipated energy in a cycle during which the parameters of a Kondo system are forced to evolve in a time-periodic manner, as in an idealized experiment. The Kondo  physics is described by the single impurity Anderson model~\cite{Anderson_PR61} (SIAM) whose Hamiltonian is 
\begin{equation}
\Ham_{\SIAM} = \Hloc + \Hhyb + \Hcond \;.
\end{equation}
The local term $\Hloc$ describes the impurity, and is given by
\begin{equation}
\Hloc = \sum_{\sigma} \varepsilon_d\, \opddag{\sigma}\opd{\sigma} + U\, \opn{\uparrow}\opn{\downarrow}+B(\opn{\uparrow}-\opn{\downarrow}) \;,
\end{equation}
with an energy level $\varepsilon_d$ and on-site Coulomb repulsion $U$. 
Additionally, we include a magnetic field $B$, whose Zeeman coupling shifts the energy levels of spin-up and spin-down in opposite directions. 
As is well known, the magnetic field destroys the Kondo effect~\cite{Hewson_kondo:book, PRL_Costi_2000}, and hence provides a control parameter that helps us extract the contribution of the Kondo effect. The impurity couples to conduction electrons through the hybridization interaction
\begin{equation}
\Hhyb = \sum_{\sigma} \sum_{k} V_{k} \, \Big( \opddag{\sigma} \, \opc{k\sigma} + \opcdag{k\sigma} \, \opd{\sigma} \Big) \;, \label{eq:Hhyb}
\end{equation}
with spin-independent hybridization couplings $V_k$. The conduction electrons are, as usual, modeled as a bath of free fermions:
\begin{align}
\Hcond = \sum_{\sigma}\sum_{k} \epsilon_k\, \opcdag{k\sigma} \, \opc{k\sigma} \;.
\end{align}

Without magnetic field, $B=0$, the model is symmetric under a spin-flip, and obeys a particle-hole (PH) symmetry if $\varepsilon_d=-U/2, V_k=V_{-k}, \epsilon_k=-\epsilon_{-k}$.  
The $B$-field breaks both spin-symmetry and PH symmetry, if present. However, at $B\neq 0$ the model is still invariant under the combined PH and spin-flip transformation
\begin{align}
\opc{k,\sigma} \longrightarrow \opcdag{-k,-\sigma} \hspace{20mm} 
\opd{\sigma} \longrightarrow -\opddag{-\sigma}\;,
\end{align}
provided we make the particle-hole symmetric choice for $U$, $V_k$ and $\epsilon_k$. 
Throughout this paper we will choose particle-hole symmetric hybridization energy dependence with a semi-circular shape, given by $V^2(x)=\Gamma \sqrt{1-x^2}/\pi W$ in the continuum limit, with hybridization coupling $\Gamma$, half-bandwidth $W$, and dimensionless energy $x=\epsilon/W$. 
Therefore, whereas at zero field the particle-hole symmetry implies $\langle \opn{\sigma}\rangle=1/2$ for equilibrium states, in presence of $B$ we have the lower symmetry $\langle \opn{\uparrow}+\opn{\downarrow}\rangle=1$ for the impurity population.

\subsection{Cyclic dissipation protocol}
We consider a time-periodic 
modulation of the Hamiltonian, to be specified case by case,  controlled by a time-dependent control parameter $\lambda$, which we assume to follow the time-dependence sketched in \cref{fig:ProtocolAndDissipation}(a).
Hence, a full cycle consists of the following four steps:

\begin{figure}[t]
	\centering
	\includegraphics[width=8.5cm]{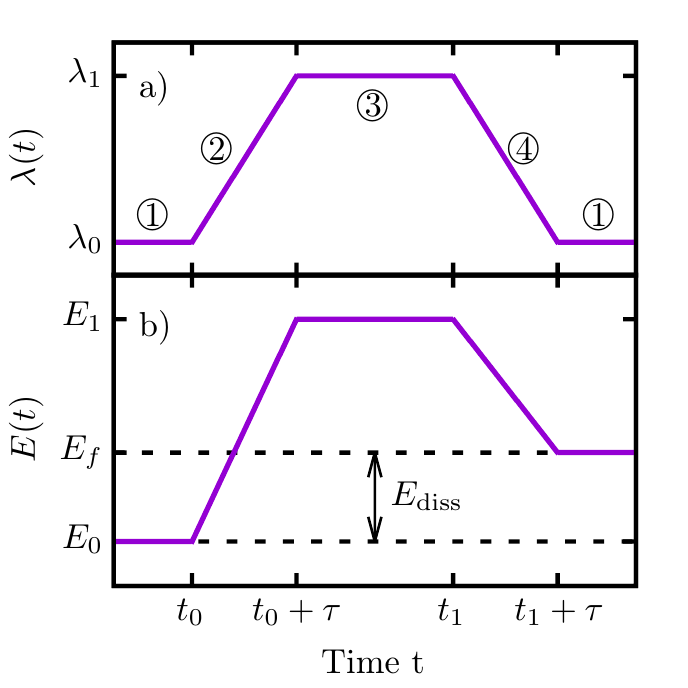}
	\caption{a) Time dependence of the driven Hamiltonian parameter $\lambda$ during a single cycle. 
		In steps \step{1} and \step{3} the Hamiltonian is kept constant and the system equilibrates. In step \step{2}, $\lambda(t)$ changes from $\lambda(t_0)\equiv\lambda_0$ to $\lambda(t_0+\tau)=\lambda_1$ in a time $\tau$, 
		with linear time-dependence. In the last step, \step{4}, $\lambda(t)$ is brought back from $\lambda_1$ to its original value $\lambda_0$. 
		b) Schematic time evolution of the total energy of the system. 
		The dissipation $E_\mathrm{diss}$ is given by the energy difference at the beginning and the end of the cycle, and is absorbed by the macroscopic bath.} 
	\label{fig:ProtocolAndDissipation}
\end{figure}

\begin{itemize}
	\item[\step{1}] The initial Hamiltonian with control parameter $\lambda(t)=\lambda_0$ is kept constant until the system has reached its equilibrium state. 
	This stage describes the tip far away from the impurity. 
	\item[\step{2}] The control parameter is raised (or lowered) linearly from $\lambda_0$ to $\lambda_1$ within a time $\tau$, modeling the transient during which the tip approaches and disturbs the impurity.
	\item[\step{3}] We let the system relax to its new equilibrium state. That is justified by a relaxation time in this problem expected to be of order $\hbar(k_B T_K)^{-1}$, typically much shorter than the time $\tau = (a/A)(2 \pi \nu)^{-1}$ during which the tip sweeps near the impurity ($h$ is Planck's constant, $a$ the impurity lateral size, $A$ the tip swing amplitude, $\nu$ the horizontal oscillation frequency).
	\item[\step{4}] In the last step, $\lambda$ turns back to its original value $\lambda_0$ of step \step{1}.
\end{itemize}
In \cref{fig:ProtocolAndDissipation}(b) we schematically show the total energy $E(t)$ of the system within a single cycle. 
During the equilibration steps \step{1} and \step{3}, the total energy is constant, as the Hamiltonian is time-independent. 
In steps \step{2} and \step{4}, owing to the time-dependence of $\lambda=\lambda(t)$, energy is forced to change and some can be pumped into the system via the impurity -- energy can flow from the impurity to or from the bath due to the hybridization coupling. 
The energy dissipation per cycle, $E_\mathrm{diss}$ is defined as the net energy pumped into the system. 
It can be calculated as the difference between $E_0=E(t=t_0)$ and $E_f=E(t=t_1+\tau)$ at the beginning and at the end of the cycle:
\begin{align}
E_{\mathrm{diss}}=E_f-E_0 \;.
\end{align}
One might ask where does the energy go, and how a closed system could continuously absorb energy. 
The answer lies in the macroscopic (infinite) size of the conduction electron bath, able to absorb a single site energy without ever heating up. 
This is particularly easy to exemplify if the bath is represented as a tight-binding chain, with energy being pumped into the impurity, and out of it into the bath. Once in the bath, energy can flow away along the chain. 
Since the chain is infinitely long in the thermodynamic limit, energy never comes back, hence it is lost forever~\cite{JMath_Chin_2010}.\\

Some more practical aspects of the simulation protocol. 
In steps \step{1} and \step{3}, one should theoretically evolve the system until it reaches equilibrium. 
Because the equilibrium state does not depend on the preceding dynamics, we can carry out steps \step{1} + \step{2} and steps \step{3} + \step{4} in separate simulations. 
The dissipation is then calculated as
\begin{align}
\begin{split}
E_{\mathrm{diss}} = \big(E(t_0+\tau)-E_0 \big)  
+ \big(E_f-E(t_1)\big) \;.
\end{split} \label{eq:dissipation}
\end{align}
Notice that this reduces to the original definition if $E(t_0+\tau)=E(t_1)$, as is the case when all steps are done in a single run, but is not true if the dynamics is split into two parts. 

\subsection{Quantum evolution method}

We compute the dissipation by simulating the full quantum dynamics of the 
time-dependent Anderson model at zero temperature. 
In this section, we briefly discuss the mathematical transformations and the technical details to carry out the simulations via matrix product states (MPS). For more details and the extension to finite temperatures via the so-called {\em thermofield transformation}~\cite{ColPhen_Umezawa_1975, PRA_Vega_2015, PRL_Weichselbaum_2018, PRB_Plenio_2020} we refer the reader to Refs.~\cite{PRB_Kohn_2021, arXiv_Kohn_2021} and references therein.

The conduction electron bath can be represented in essentially two geometries: (i) The star geometry mimics the geometry of the interactions, see \cref{eq:Hhyb}, where any conduction electron mode interacts with the impurity. In MPS simulations, this geometry requires dealing with long-range interactions, but benefits from very low entanglement~\cite{PRB_Wolf_2014}. The second possible geometry is the chain geometry, where the conduction electrons are mapped into a nearest-neighbor chain~\cite{Wilson_RMP75,RMP_Bulla_2008}. This geometry is suitable for tensor network methods due to the interactions being only of nearest-neighbor distance, but it suffers from larger entanglement. In this paper, we employ an improved chain mapping of the conduction electrons, with {\it both} short-range interactions and low entanglement~\cite{PRB_Kohn_2021}. The essential idea of the technique is a separation of electron modes above and below the bath chemical potential, followed by an independent chain mapping. 
In this way, we avoid the detrimental mixing of filled and empty modes, and we preserve the product state property of the conduction bath's ground state. Mathematically we define two fermionic operators 

\begin{align}
    \opa{1,0,\sigma} = J_{1,0}^{-1} \sum_{k, \epsilon_k>\mu } V_k \, \opc{k,\sigma} \\
    \opa{2,0,\sigma} = J_{2,0}^{-1} \sum_{k, \epsilon_k\leq \mu} V_k \, \opc{k,\sigma},
\end{align}
where $J_{c,0}$ ($c=1,2$) ensures correct normalization $\{\opa{c,0,\sigma},\opadag{c',0,\sigma'} \} = \delta_{c,c'}\delta_{\sigma,\sigma'}$.
The hybridization term then becomes
\begin{align}
    \Hhyb = \sum_{\sigma}\sum_{c=1}^{2} J_{c,0}\left( \opddag{\sigma}\opa{c,0,\sigma} + \opadag{c,0,\sigma}\opd{\sigma} \right) \;.
\end{align}

Similar to the original chain mapping~\cite{Wilson_RMP75,RMP_Bulla_2008}, we can apply Lanczos' algorithm independently to $\opa{1,0,\sigma}$ and $\opa{2,0,\sigma}$, to obtain two non-interacting chains with fermionic operators $\opa{c,n,\sigma}$.

In the electron bath ground state, modes with energy above (below) the chemical potential $\mu$ are empty (filled). Since  $\opa{1,n,\sigma}$ and $\opa{2,n,\sigma}$ are linear combinations of modes that are empty (filled) in the ground state, they are completely empty (filled) as well, and hence, the conduction electron bath ground state is a simple product state, which, as demonstrated in Refs.~\cite{PRB_Kohn_2021, arXiv_Kohn_2021}, is highly beneficial for the simulations. The final Hamiltonian to simulate consists of two Wilson's chains, and the impurity interacting with the first site of both chains:

\begin{widetext}
\begin{equation}
\Ham_{\SIAM}(t) = \Hloc(t)  +  \sum_{\sigma} \sum_{c=1}^2 \Bigg( J_{\chain,0} \left(\opddag{\sigma} \, \opa{\chain,0,\sigma} + \Hc \right) 
+  \sum_{n=0}  \Big( E_{\chain,n} \, \opadag{\chain,n,\sigma}\opa{\chain,n,\sigma} + \big(J_{\chain,n+1} \opadag{\chain,n+1,\sigma}\opa{\chain,n,\sigma} + \Hc \big) \Big) \Bigg) \;.
\end{equation}
\end{widetext}
Here, the chain coefficients $E_{\chain,n}$ and $J_{\chain,n}$ are obtained from the chain mapping.
The one dimensional structure of $\Ham_{\SIAM}$ together with the short-range nature of interactions allows us to carry out simulations efficiently using matrix product states~\cite{Schollwoeck_RMP05,AoP_Schollwoeck_2011}. 
For equilibrium simulations we employ the density matrix renormalization group (DMRG) algorithm to find ground states~\cite{White_PRL92}. 
The real-time dynamics of the system with time-dependent Hamiltonian is computed using the recently developed time-dependent variational principle (TDVP) algorithm in its two-site variant~\cite{PRB_Haegeman_2016}, where two neighboring tensors of the MPS are evolved in time together in each step. This algorithm has proven to deliver very accurate results at low computational costs~\cite{AoP_Paeckel_2019}.

\begin{figure*}[ht]
	\centering
	\includegraphics[width=15cm]{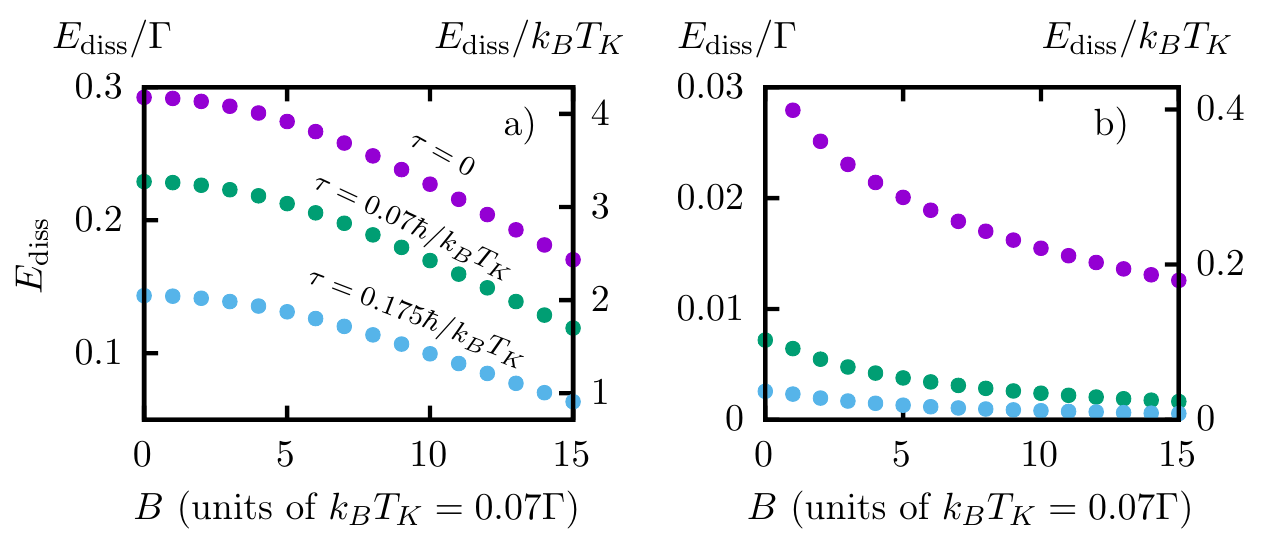}
	\caption{Dissipation per cycle in (a) the noninteracting $U=0$ and (b) the interacting case $U=2.5\pi\Gamma$ at zero temperature, for different ramp times $\tau$, as a function of the static magnetic field. 
	The impurity level is driven between the particle-hole symmetric value $\varepsilon_0=-U/2$ and $\varepsilon_1=\varepsilon_0+\Delta\varepsilon$, with $\Delta\varepsilon=0.7\Gamma=10k_BT_K$ where the Kondo temperature in the interacting case is $k_BT_K=0.07\Gamma$. 
	Dissipation is given in units of $\Gamma$ (left scale) and in units of the Kondo temperature (right scale).}
	\label{fig:DissipationT0NonintDeltaEps+10kbTk}
\end{figure*}

At zero temperature, there is one more simplification possible. Instead of calculating the equilibrium state through a preliminary real-time annealing evolution~\cite{PRB_Kohn_2021}, we can simply use the DMRG algorithm to calculate the ground state of the system variationally. Even if finite temperature simulations are possible as well~\cite{PRB_Kohn_2021}, here we will for simplicity restrict ourselves to zero temperature, where the study of switching-time dependence is more directly addressed.

\section{Results} \label{sec:results}
This section describes our results for the dissipation calculated at zero temperature, during a cycle where the forcing perturbation varies as in Fig.1. 
They should tell us whether or not the dissipation shows signatures of the 
forced switching of the Kondo effect with two different types of time-periodic cyclic forcing. As anticipated, one consists of a variation of the SIAM impurity energy level periodically up and down. across the Fermi level. 
The second involves the application of a cyclic magnetic field, varying from zero to a value sufficient to destroy the Kondo effect, and back.

\subsection{Time-dependent impurity energy level $\varepsilon_d$}
Here we consider (without or with a static magnetic field $B$) a time-dependent on-site energy level, taking  $\lambda(t)=\varepsilon_d(t)$. 

With the protocol discussed above, the impurity energy level is periodically driven between $\varepsilon_0$ and $\varepsilon_1$, where we choose the particle-hole symmetric situation $\varepsilon_0=-U/2$ for the initial state, and $\varepsilon_1=-U/2+0.7\Gamma$ for the intermediate step \step{3}.
The Kondo temperature for $U=2.5\pi\Gamma$ and at PH-symmetry is $k_BT_K=0.07\Gamma$~\cite{PRB_Kohn_2021}, and hence the energy level is shifted by $\Delta\varepsilon=\varepsilon_1-\varepsilon_0=0.7\Gamma=10k_BT_K$.

\begin{figure*}[ht]
	\centering
	\includegraphics[width=15cm]{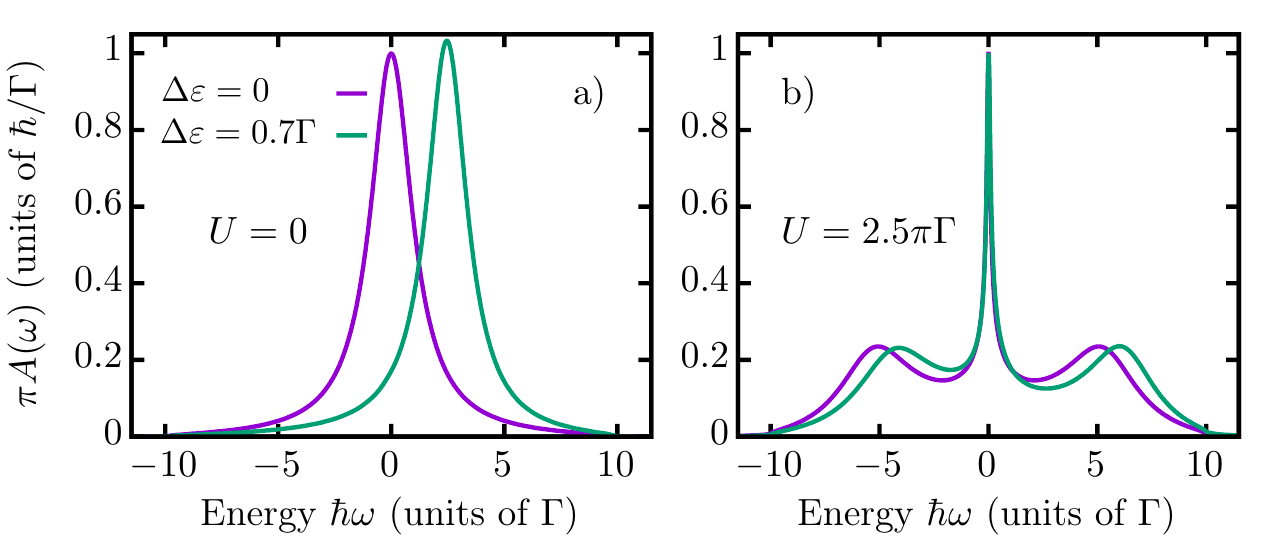}
	\caption{Zero temperature equilibrium spectral function in the noninterating $U=0$ (a) and interacting case $U=2.5\pi\Gamma$ (b), the latter with Kondo temperature $k_BT_K=0.07\Gamma$. 
	Two different choices of the impurity energy level $\epsilon_d=-U/2+\Delta\varepsilon$ are considered: the particle-hole symmetric choice $\Delta\varepsilon=0$, and an asymmetric one with $\Delta\varepsilon=0.7\Gamma$.}
	\label{fig:Spectral_functions}
\end{figure*}

We start by considering the noninteracting case, $U=0$. 
Here, the impurity spectral function has just a single peak (a Friedel resonance), corresponding to the local level of the impurity, broadened due to the hybridization coupling, see \cref{fig:Spectral_functions}(a). 
For a sudden quench ($\tau=0$) and zero magnetic field $B=0$, we find the dissipation $E_\mathrm{diss}=0.29\Gamma$ to be in the order of the level shift $\Delta\varepsilon=0.7\Gamma$.
The dissipation decreases monotonically with increasing magnetic field, see \cref{fig:DissipationT0NonintDeltaEps+10kbTk}(a), presumably due to the opposite and compensating effect of the field on the spin-up and spin-down impurity levels. Furthermore, dissipation is reduced by slowing down the time-dependent cycle, 
as expected, since dissipation must vanish in the limit of an adiabatic evolution.

Let us move next to the interacting case with $U=2.5\pi\Gamma$, and corresponding Kondo temperature $k_BT_K=0.07\Gamma$. 
Again, the impurity level energy is lifted by $\Delta\varepsilon=0.07\Gamma=10k_BT_K$. 
The dissipation turns out to be significantly lower as compared to the noninteracting case, by about one order of magnitude even for the sudden quench, 
see \cref{fig:DissipationT0NonintDeltaEps+10kbTk}. 

As an incidental note, 
the magnetic field dependence of dissipation at $\tau=0$ follows the (inverse) behavior of the impurity population at $\varepsilon_1$: The occupation is field-independent at 
particle-hole symmetry $\varepsilon_d=\varepsilon_0$
while dissipation is given by the difference in equilibrium occupations at $\varepsilon_0$ and $\varepsilon_1$, which follows immediately from \cref{eq:dissipation} for a sudden quench of the impurity energy level $\varepsilon_d$.
The drop of dissipation with increasing ramp time turns out to be very rapid. In fact, the time scale on which dissipation disappears is clearly smaller than the Kondo time scale $\hbar/k_BT_K$, in contrast to what one would expect for the case of dissipation emerging from the Kondo effect. This issue will be discussed in the next section.

\begin{figure}[ht]
	\centering
	\includegraphics[width=8cm]{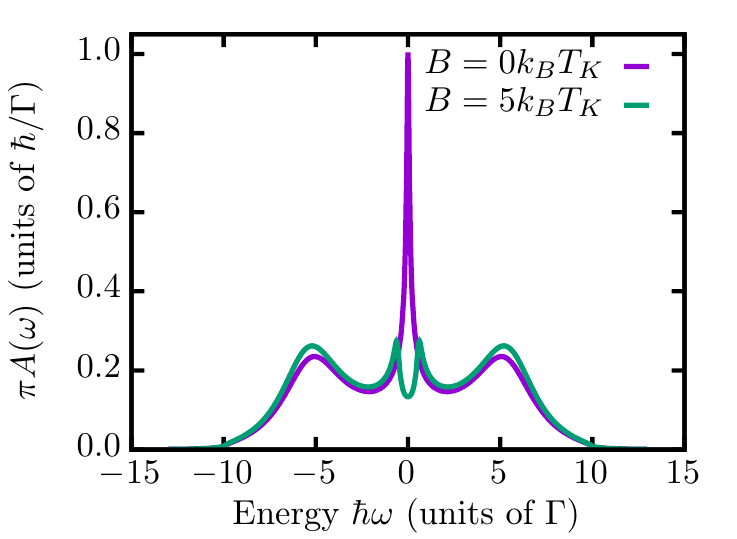}
	\caption{Spectral function of the SIAM in absence of a magnetic field, $B=0$, and for $B=5k_BT_K$. The model is particle-hole symmetric with $U=2.5\pi\Gamma$, corresponding to a Kondo temperature of $k_BT_K=0.07\Gamma$. The spectral function was calculated with the method presented in Ref.~\cite{PRB_Kohn_2021} and averaged over spin-up and spin-down, see e.g. Refs.~\cite{NJP_vonDelft_2018,NJP_Zitko_2010}. }
	\label{fig:SpectralFctDifB}
\end{figure}

To get a better understanding of the mechanisms leading to the observed dissipation, let us analyze the equilibrium impurity spectral function $A(\omega)$, obtained from MPS calculations, 
as discussed in more detail in Ref.~\cite{PRB_Kohn_2021}.
The noninteracting $U=0$ spectral function has a single peak, due to the local impurity level, see \cref{fig:Spectral_functions}(a). 
The interacting spectral function for $U=2.5\pi\Gamma$, see \cref{fig:Spectral_functions}(b), instead, shows two peaks corresponding to the local impurity levels at $\hbar\omega=\varepsilon_d$ and 
$\hbar\omega=\varepsilon_d+U$, and a Kondo peak at the conduction electron Fermi energy, here set to $E_F=0$. 
By shifting the impurity energy level from $\Delta\varepsilon=0$ to $\Delta\varepsilon=10k_BT_K$ the two side-peaks move accordingly. 
However, the Kondo peak is barely affected by this 
impurity level shift. 
The overall change of the spectral function upon moving the impurity level is much more significant in the noninteracting case, which 
is clearly the cause of
the much larger dissipation, yet not of Kondo origin. 
Summing up this warm-up exercise, 
a time-dependent chemical potential oscillation shifting the impurity level 
will not give rise to Kondo dissipation. 
The physical reason for this, as shown by \cref{fig:Spectral_functions}(b), is that the Kondo peak is barely affected by the perturbation.

In order to cause the Kondo switching electrostatically, the impurity level should be switched in a rather drastical manner, say from well below $E_F-U$ (doubly occupied impurity, no Kondo) to near $E_F-U/2$ (a singly occupied impurity, Kondo regime), or from the latter to well above $E_F+U$ (empty impurity, no Kondo). 
Another perturbation that will lead to Kondo dissipation is, as implemented in Ref. \cite{PRB_Baruselli_2017}, a periodic on-off switching of impurity-bath hybridization. 
However, both such extreme perturbations will in real life be accompanied by an unpredictably large amount of subsidiary dissipation of non-Kondo origin. 
Therefore, instead of analysing further these cases, we move directly to the -- presumably less dramatic -- magnetic switching.  

\subsection{Time-dependent magnetic field}
We just saw that a gentle electron impurity level switching fails to produce Kondo dissipation, because it leaves the narrow Kondo resonance unchanged.
On the other hand, an external magnetic field is well known to quench and split the Kondo peak.~\cite{PRL_Costi_2000, NJP_vonDelft_2018}

We study the symmetric SIAM ($\varepsilon_d=-U/2$) with a time dependent magnetic field $B(t)$, where $B$ changes between $B_0=B(t=t_0)=0$ and $B_1=B(t_0+\tau)=5k_BT_K$. 

\begin{figure}[ht]
\centering
\includegraphics[width=8cm]{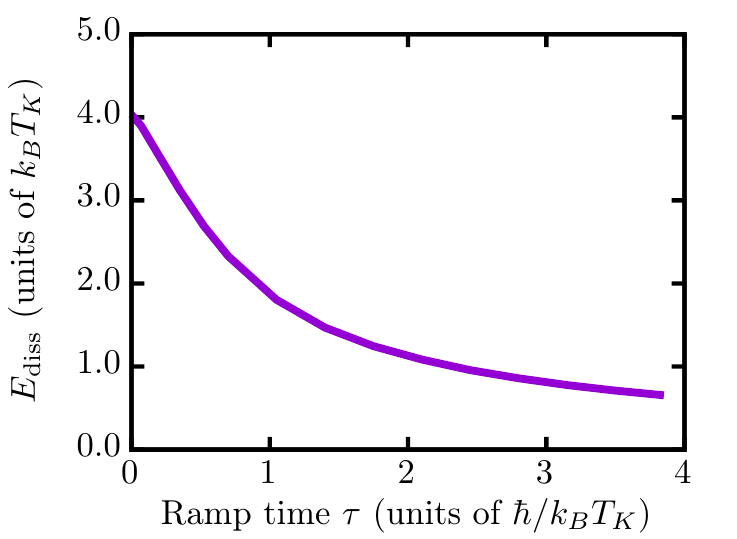}
\caption{Dissipation per cycle for the protocol where the magnetic field is changed linearly changed between $B_0=0$ to $B_1=5k_BT_K$. The dissipation is in the order of the Kondo temperature and decays on a time scale $t\propto \hbar(k_B T_K)^{-1}$.}
\label{fig:DissTimeDepB}
\end{figure}

Let us  consider in detail the (spin-averaged) spectral function in presence of a static magnetic field. 
As shown in \cref{fig:SpectralFctDifB}, the Kondo peak is essentially gone already at $B=5k_BT_K$,  making the protocol with increasing magnetic field up to $B_1=5k_BT_K$ a promising candidate to observe Kondo related dissipation. Importantly, the local peaks are barely modified by the magnetic field, and hence we expect only a minor contribution to the dissipation.

Once we make the field $B$ time-dependent with the standard on-off protocol, see \cref{fig:ProtocolAndDissipation}, the dissipation associated with sudden switching of $B$ is indeed significantly larger, in the order of $E_{\mathrm{diss}}=4k_BT_K$, as compared to the small dissipation for a time-dependent on-site energy level $\varepsilon_d$. Moreover, as shown by the final result of \cref{fig:DissTimeDepB}, the dissipation decays much slower once the the ramp time $\tau$ is progressively increased from zero.
In fact, the dissipation decays on the time scale of the inverse Kondo temperature, $\hbar(k_B T_K)^{-1}$. A clear rationale is that field variations on larger time scales are felt as essentially adiabatic, and hence, cause negligible dissipation. We expect that the very same behavior applies to asymmetric switches of the impurity level energy, and to the switching of the hybridization energy considered by Baruselli \etal ~\cite{PRB_Baruselli_2017}.

\section{Discussion and conclusions} \label{sec:conclusions}

We have shown that suppression of the Kondo effect by a properly switched magnetic will cause the expected dissipation. A large dissipation, in the order of $E_{\mathrm{diss}}=4k_BT_K$ per cycle, equal to that originally predicted by Baruselli {\em et al.} \cite{PRB_Baruselli_2017}, will occur if the Kondo switch-off time is sufficiently short, typically $\hbar (k_B T_K)^{-1}$ or shorter.
On the contrary, a much smaller dissipation --- mostly related to non-Kondo sideband effects in the spectral density ---, is seen when the impurity level occcupation is shifted by similarly large values of the order of $\Delta \epsilon \sim 10 k_BT_K$. 
Obviously, extremely large oscillations, completely destroying the Kondo peak, are likely to present more dissipation, but still, it would be hard to discriminate the contribution purely due to the Kondo effect from the large sideband electronic effects. 
%
Technically, this work represents a nontrivial application of a matrix-product-state based, time-dependent variational algorithm freshly established by some of us~\cite{PRB_Kohn_2021, arXiv_Kohn_2021}. 

Our main result is that Kondo dissipation drops very fast with switching time. As Fig.~\ref{fig:DissTimeDepB} shows, the dissipation has dropped to about 1/4 of its value once the switching time grows from zero -- the sudden switch limit of Ref.~\cite{PRB_Baruselli_2017} -- to about $2.5\hbar(k_B T_K)^{-1}$.  
While this result makes good physical sense, it does pose an experimental problem for the possibility that the Kondo switching dissipation could be observed by e.g., a non-contact pendulum AFM. 
Consider a tip flying above the surface-deposited Kondo impurity of size $a$ with frequency $\nu$, usually not larger than tens of kHz, and large amplitude $A$, larger that the atomic impurity size $a\sim 0.2$ nm, but ordinarily below 10 nm.  
In the most optimistic case, the tip sway time $a/(A\nu)$ over the impurity, during which Kondo could be switched off and on, could be shrunk down to perhaps a microsecond, still orders of magnitude longer than $\hbar(k_B T_K)^{-1}$, a time lasting at most tens of picoseconds.

In conclusion, cyclic switching of a magnetic impurity from Kondo to a non-Kondo state is predicted to involve a very important dissipation of the order of several $k_BT_K$ per cycle. That dissipation critically depends on a sufficiently fast switching time, typically the Kondo time $\hbar(k_B T_K)^{-1}$ or faster. While such a fast switching seems problematic for standard AFM setups, the challenge remains open for other possible means to detect this dissipation by time dependent magnetic field, electrostatic impurity level shift, or hybridization switching.

\section*{Acknowledgments}
This work was carried out under European Unions H2020 Framework Programme/ERC Advanced Grant N. 8344023 ULTRADISS. 
GES acknowledges that his research has been conducted within the framework of the Trieste Institute for Theoretical Quantum Technologies (TQT).
We thank Rémy Pawlak and Ernst Meyer (University of Basel) for many helpful discussions. Simulations were performed using the ITensor library \cite{itensor}.

\bibliographystyle{apsrev4-1}
\bibliography{BiblioMPS_OQS}

\begin{thebibliography}{34}%
\makeatletter
\providecommand \@ifxundefined [1]{%
 \@ifx{#1\undefined}
}%
\providecommand \@ifnum [1]{%
 \ifnum #1\expandafter \@firstoftwo
 \else \expandafter \@secondoftwo
 \fi
}%
\providecommand \@ifx [1]{%
 \ifx #1\expandafter \@firstoftwo
 \else \expandafter \@secondoftwo
 \fi
}%
\providecommand \natexlab [1]{#1}%
\providecommand \enquote  [1]{``#1''}%
\providecommand \bibnamefont  [1]{#1}%
\providecommand \bibfnamefont [1]{#1}%
\providecommand \citenamefont [1]{#1}%
\providecommand \href@noop [0]{\@secondoftwo}%
\providecommand \href [0]{\begingroup \@sanitize@url \@href}%
\providecommand \@href[1]{\@@startlink{#1}\@@href}%
\providecommand \@@href[1]{\endgroup#1\@@endlink}%
\providecommand \@sanitize@url [0]{\catcode `\\12\catcode `\$12\catcode
  `\&12\catcode `\#12\catcode `\^12\catcode `\_12\catcode `\%12\relax}%
\providecommand \@@startlink[1]{}%
\providecommand \@@endlink[0]{}%
\providecommand \url  [0]{\begingroup\@sanitize@url \@url }%
\providecommand \@url [1]{\endgroup\@href {#1}{\urlprefix }}%
\providecommand \urlprefix  [0]{URL }%
\providecommand \Eprint [0]{\href }%
\providecommand \doibase [0]{http://dx.doi.org/}%
\providecommand \selectlanguage [0]{\@gobble}%
\providecommand \bibinfo  [0]{\@secondoftwo}%
\providecommand \bibfield  [0]{\@secondoftwo}%
\providecommand \translation [1]{[#1]}%
\providecommand \BibitemOpen [0]{}%
\providecommand \bibitemStop [0]{}%
\providecommand \bibitemNoStop [0]{.\EOS\space}%
\providecommand \EOS [0]{\spacefactor3000\relax}%
\providecommand \BibitemShut  [1]{\csname bibitem#1\endcsname}%
\let\auto@bib@innerbib\@empty
\bibitem [{\citenamefont {Anderson}\ and\ \citenamefont
  {Yuval}(1969)}]{PRL_Anderson_1969}%
  \BibitemOpen
  \bibfield  {author} {\bibinfo {author} {\bibfnamefont {P.~W.}\ \bibnamefont
  {Anderson}}\ and\ \bibinfo {author} {\bibfnamefont {G.}~\bibnamefont
  {Yuval}},\ }\href {\doibase 10.1103/PhysRevLett.23.89} {\bibfield  {journal}
  {\bibinfo  {journal} {Phys. Rev. Lett.}\ }\textbf {\bibinfo {volume} {23}},\
  \bibinfo {pages} {89} (\bibinfo {year} {1969})}\BibitemShut {NoStop}%
\bibitem [{\citenamefont {Anderson}\ \emph {et~al.}(1970)\citenamefont
  {Anderson}, \citenamefont {Yuval},\ and\ \citenamefont
  {Hamann}}]{PRB_Anderson_1970}%
  \BibitemOpen
  \bibfield  {author} {\bibinfo {author} {\bibfnamefont {P.~W.}\ \bibnamefont
  {Anderson}}, \bibinfo {author} {\bibfnamefont {G.}~\bibnamefont {Yuval}}, \
  and\ \bibinfo {author} {\bibfnamefont {D.~R.}\ \bibnamefont {Hamann}},\
  }\href {\doibase 10.1103/PhysRevB.1.4464} {\bibfield  {journal} {\bibinfo
  {journal} {Phys. Rev. B}\ }\textbf {\bibinfo {volume} {1}},\ \bibinfo {pages}
  {4464} (\bibinfo {year} {1970})}\BibitemShut {NoStop}%
\bibitem [{\citenamefont {Appelbaum}(1967)}]{PR_Appelbaum_1967}%
  \BibitemOpen
  \bibfield  {author} {\bibinfo {author} {\bibfnamefont {J.~A.}\ \bibnamefont
  {Appelbaum}},\ }\href {\doibase 10.1103/PhysRev.154.633} {\bibfield
  {journal} {\bibinfo  {journal} {Phys. Rev.}\ }\textbf {\bibinfo {volume}
  {154}},\ \bibinfo {pages} {633} (\bibinfo {year} {1967})}\BibitemShut
  {NoStop}%
\bibitem [{\citenamefont {Cockins}\ \emph {et~al.}(2010)\citenamefont
  {Cockins}, \citenamefont {Miyahara}, \citenamefont {Bennett}, \citenamefont
  {Clerk}, \citenamefont {Studenikin}, \citenamefont {Poole}, \citenamefont
  {Sachrajda},\ and\ \citenamefont {Grutter}}]{PNAS_Cockins_2010}%
  \BibitemOpen
  \bibfield  {author} {\bibinfo {author} {\bibfnamefont {L.}~\bibnamefont
  {Cockins}}, \bibinfo {author} {\bibfnamefont {Y.}~\bibnamefont {Miyahara}},
  \bibinfo {author} {\bibfnamefont {S.~D.}\ \bibnamefont {Bennett}}, \bibinfo
  {author} {\bibfnamefont {A.~A.}\ \bibnamefont {Clerk}}, \bibinfo {author}
  {\bibfnamefont {S.}~\bibnamefont {Studenikin}}, \bibinfo {author}
  {\bibfnamefont {P.}~\bibnamefont {Poole}}, \bibinfo {author} {\bibfnamefont
  {A.}~\bibnamefont {Sachrajda}}, \ and\ \bibinfo {author} {\bibfnamefont
  {P.}~\bibnamefont {Grutter}},\ }\href {\doibase 10.1073/pnas.0912716107}
  {\bibfield  {journal} {\bibinfo  {journal} {Proceedings of the National
  Academy of Sciences}\ }\textbf {\bibinfo {volume} {107}},\ \bibinfo {pages}
  {9496} (\bibinfo {year} {2010})}\BibitemShut {NoStop}%
\bibitem [{\citenamefont {Kisiel}\ \emph {et~al.}(2018)\citenamefont {Kisiel},
  \citenamefont {Brovko}, \citenamefont {Yildiz}, \citenamefont {Pawlak},
  \citenamefont {Gysin}, \citenamefont {Tosatti},\ and\ \citenamefont
  {Meyer}}]{Nature_Kisiel_2018}%
  \BibitemOpen
  \bibfield  {author} {\bibinfo {author} {\bibfnamefont {M.}~\bibnamefont
  {Kisiel}}, \bibinfo {author} {\bibfnamefont {O.~O.}\ \bibnamefont {Brovko}},
  \bibinfo {author} {\bibfnamefont {D.}~\bibnamefont {Yildiz}}, \bibinfo
  {author} {\bibfnamefont {R.}~\bibnamefont {Pawlak}}, \bibinfo {author}
  {\bibfnamefont {U.}~\bibnamefont {Gysin}}, \bibinfo {author} {\bibfnamefont
  {E.}~\bibnamefont {Tosatti}}, \ and\ \bibinfo {author} {\bibfnamefont
  {E.}~\bibnamefont {Meyer}},\ }\href {\doibase 10.1038/s41467-018-05392-1}
  {\bibfield  {journal} {\bibinfo  {journal} {Nature Communications}\ }\textbf
  {\bibinfo {volume} {9}},\ \bibinfo {pages} {2946} (\bibinfo {year}
  {2018})}\BibitemShut {NoStop}%
\bibitem [{\citenamefont {Gotsmann}\ \emph {et~al.}(1999)\citenamefont
  {Gotsmann}, \citenamefont {Seidel}, \citenamefont {Anczykowski},\ and\
  \citenamefont {Fuchs}}]{PRB_Fuchs_1999}%
  \BibitemOpen
  \bibfield  {author} {\bibinfo {author} {\bibfnamefont {B.}~\bibnamefont
  {Gotsmann}}, \bibinfo {author} {\bibfnamefont {C.}~\bibnamefont {Seidel}},
  \bibinfo {author} {\bibfnamefont {B.}~\bibnamefont {Anczykowski}}, \ and\
  \bibinfo {author} {\bibfnamefont {H.}~\bibnamefont {Fuchs}},\ }\href
  {\doibase 10.1103/PhysRevB.60.11051} {\bibfield  {journal} {\bibinfo
  {journal} {Phys. Rev. B}\ }\textbf {\bibinfo {volume} {60}},\ \bibinfo
  {pages} {11051} (\bibinfo {year} {1999})}\BibitemShut {NoStop}%
\bibitem [{\citenamefont {Kisiel}\ \emph {et~al.}(2011)\citenamefont {Kisiel},
  \citenamefont {Gnecco}, \citenamefont {Gysin}, \citenamefont {Marot},
  \citenamefont {Rast},\ and\ \citenamefont {Meyer}}]{Nature_Kisiel_2011}%
  \BibitemOpen
  \bibfield  {author} {\bibinfo {author} {\bibfnamefont {M.}~\bibnamefont
  {Kisiel}}, \bibinfo {author} {\bibfnamefont {E.}~\bibnamefont {Gnecco}},
  \bibinfo {author} {\bibfnamefont {U.}~\bibnamefont {Gysin}}, \bibinfo
  {author} {\bibfnamefont {L.}~\bibnamefont {Marot}}, \bibinfo {author}
  {\bibfnamefont {S.}~\bibnamefont {Rast}}, \ and\ \bibinfo {author}
  {\bibfnamefont {E.}~\bibnamefont {Meyer}},\ }\href {\doibase
  10.1038/nmat2936} {\bibfield  {journal} {\bibinfo  {journal} {Nature
  Materials}\ }\textbf {\bibinfo {volume} {10}},\ \bibinfo {pages} {119}
  (\bibinfo {year} {2011})}\BibitemShut {NoStop}%
\bibitem [{\citenamefont {Langer}\ \emph {et~al.}(2014)\citenamefont {Langer},
  \citenamefont {Kisiel}, \citenamefont {Pawlak}, \citenamefont {Pellegrini},
  \citenamefont {Santoro}, \citenamefont {Buzio}, \citenamefont {Gerbi},
  \citenamefont {Balakrishnan}, \citenamefont {Baratoff}, \citenamefont
  {Tosatti},\ and\ \citenamefont {Meyer}}]{Nature_Langer_2014}%
  \BibitemOpen
  \bibfield  {author} {\bibinfo {author} {\bibfnamefont {M.}~\bibnamefont
  {Langer}}, \bibinfo {author} {\bibfnamefont {M.}~\bibnamefont {Kisiel}},
  \bibinfo {author} {\bibfnamefont {R.}~\bibnamefont {Pawlak}}, \bibinfo
  {author} {\bibfnamefont {F.}~\bibnamefont {Pellegrini}}, \bibinfo {author}
  {\bibfnamefont {G.~E.}\ \bibnamefont {Santoro}}, \bibinfo {author}
  {\bibfnamefont {R.}~\bibnamefont {Buzio}}, \bibinfo {author} {\bibfnamefont
  {A.}~\bibnamefont {Gerbi}}, \bibinfo {author} {\bibfnamefont
  {G.}~\bibnamefont {Balakrishnan}}, \bibinfo {author} {\bibfnamefont
  {A.}~\bibnamefont {Baratoff}}, \bibinfo {author} {\bibfnamefont
  {E.}~\bibnamefont {Tosatti}}, \ and\ \bibinfo {author} {\bibfnamefont
  {E.}~\bibnamefont {Meyer}},\ }\href {\doibase 10.1038/nmat3836} {\bibfield
  {journal} {\bibinfo  {journal} {Nature Materials}\ }\textbf {\bibinfo
  {volume} {13}},\ \bibinfo {pages} {173} (\bibinfo {year} {2014})}\BibitemShut
  {NoStop}%
\bibitem [{\citenamefont {Kisiel}\ \emph {et~al.}(2015)\citenamefont {Kisiel},
  \citenamefont {Pellegrini}, \citenamefont {Santoro}, \citenamefont
  {Samadashvili}, \citenamefont {Pawlak}, \citenamefont {Benassi},
  \citenamefont {Gysin}, \citenamefont {Buzio}, \citenamefont {Gerbi},
  \citenamefont {Meyer},\ and\ \citenamefont {Tosatti}}]{PRL_Meyer_2015}%
  \BibitemOpen
  \bibfield  {author} {\bibinfo {author} {\bibfnamefont {M.}~\bibnamefont
  {Kisiel}}, \bibinfo {author} {\bibfnamefont {F.}~\bibnamefont {Pellegrini}},
  \bibinfo {author} {\bibfnamefont {G.~E.}\ \bibnamefont {Santoro}}, \bibinfo
  {author} {\bibfnamefont {M.}~\bibnamefont {Samadashvili}}, \bibinfo {author}
  {\bibfnamefont {R.}~\bibnamefont {Pawlak}}, \bibinfo {author} {\bibfnamefont
  {A.}~\bibnamefont {Benassi}}, \bibinfo {author} {\bibfnamefont
  {U.}~\bibnamefont {Gysin}}, \bibinfo {author} {\bibfnamefont
  {R.}~\bibnamefont {Buzio}}, \bibinfo {author} {\bibfnamefont
  {A.}~\bibnamefont {Gerbi}}, \bibinfo {author} {\bibfnamefont
  {E.}~\bibnamefont {Meyer}}, \ and\ \bibinfo {author} {\bibfnamefont
  {E.}~\bibnamefont {Tosatti}},\ }\href {\doibase
  10.1103/PhysRevLett.115.046101} {\bibfield  {journal} {\bibinfo  {journal}
  {Phys. Rev. Lett.}\ }\textbf {\bibinfo {volume} {115}},\ \bibinfo {pages}
  {046101} (\bibinfo {year} {2015})}\BibitemShut {NoStop}%
\bibitem [{\citenamefont {Baruselli}\ \emph {et~al.}(2017)\citenamefont
  {Baruselli}, \citenamefont {Fabrizio},\ and\ \citenamefont
  {Tosatti}}]{PRB_Baruselli_2017}%
  \BibitemOpen
  \bibfield  {author} {\bibinfo {author} {\bibfnamefont {P.~P.}\ \bibnamefont
  {Baruselli}}, \bibinfo {author} {\bibfnamefont {M.}~\bibnamefont {Fabrizio}},
  \ and\ \bibinfo {author} {\bibfnamefont {E.}~\bibnamefont {Tosatti}},\ }\href
  {\doibase 10.1103/PhysRevB.96.075113} {\bibfield  {journal} {\bibinfo
  {journal} {Phys. Rev. B}\ }\textbf {\bibinfo {volume} {96}},\ \bibinfo
  {pages} {075113} (\bibinfo {year} {2017})}\BibitemShut {NoStop}%
\bibitem [{\citenamefont {Baruselli}\ and\ \citenamefont
  {Tosatti}(2018)}]{ArXiv_Baruselli_2018}%
  \BibitemOpen
  \bibfield  {author} {\bibinfo {author} {\bibfnamefont {P.~P.}\ \bibnamefont
  {Baruselli}}\ and\ \bibinfo {author} {\bibfnamefont {E.}~\bibnamefont
  {Tosatti}},\ }\href@noop {} {\  (\bibinfo {year} {2018})},\ \Eprint
  {http://arxiv.org/abs/1804.04999} {arXiv:1804.04999} \BibitemShut {NoStop}%
\bibitem [{\citenamefont {Jacobson}\ \emph {et~al.}(2017)\citenamefont
  {Jacobson}, \citenamefont {Muenks}, \citenamefont {Laskin}, \citenamefont
  {Brovko}, \citenamefont {Stepanyuk}, \citenamefont {Ternes},\ and\
  \citenamefont {Kern}}]{Science_Jacobson_2017}%
  \BibitemOpen
  \bibfield  {author} {\bibinfo {author} {\bibfnamefont {P.}~\bibnamefont
  {Jacobson}}, \bibinfo {author} {\bibfnamefont {M.}~\bibnamefont {Muenks}},
  \bibinfo {author} {\bibfnamefont {G.}~\bibnamefont {Laskin}}, \bibinfo
  {author} {\bibfnamefont {O.}~\bibnamefont {Brovko}}, \bibinfo {author}
  {\bibfnamefont {V.}~\bibnamefont {Stepanyuk}}, \bibinfo {author}
  {\bibfnamefont {M.}~\bibnamefont {Ternes}}, \ and\ \bibinfo {author}
  {\bibfnamefont {K.}~\bibnamefont {Kern}},\ }\href {\doibase
  10.1126/sciadv.1602060} {\bibfield  {journal} {\bibinfo  {journal} {Science
  Advances}\ }\textbf {\bibinfo {volume} {3}} (\bibinfo {year} {2017}),\
  10.1126/sciadv.1602060}\BibitemShut {NoStop}%
\bibitem [{\citenamefont {Pawlak}\ \emph {et~al.}(2016)\citenamefont {Pawlak},
  \citenamefont {Ouyang}, \citenamefont {Filippov}, \citenamefont
  {Kalikhman-Razvozov}, \citenamefont {Kawai}, \citenamefont {Glatzel},
  \citenamefont {Gnecco}, \citenamefont {Baratoff}, \citenamefont {Zheng},
  \citenamefont {Hod}, \citenamefont {Urbakh},\ and\ \citenamefont
  {Meyer}}]{ACS_Pawlak_2016}%
  \BibitemOpen
  \bibfield  {author} {\bibinfo {author} {\bibfnamefont {R.}~\bibnamefont
  {Pawlak}}, \bibinfo {author} {\bibfnamefont {W.}~\bibnamefont {Ouyang}},
  \bibinfo {author} {\bibfnamefont {A.~E.}\ \bibnamefont {Filippov}}, \bibinfo
  {author} {\bibfnamefont {L.}~\bibnamefont {Kalikhman-Razvozov}}, \bibinfo
  {author} {\bibfnamefont {S.}~\bibnamefont {Kawai}}, \bibinfo {author}
  {\bibfnamefont {T.}~\bibnamefont {Glatzel}}, \bibinfo {author} {\bibfnamefont
  {E.}~\bibnamefont {Gnecco}}, \bibinfo {author} {\bibfnamefont
  {A.}~\bibnamefont {Baratoff}}, \bibinfo {author} {\bibfnamefont
  {Q.}~\bibnamefont {Zheng}}, \bibinfo {author} {\bibfnamefont
  {O.}~\bibnamefont {Hod}}, \bibinfo {author} {\bibfnamefont {M.}~\bibnamefont
  {Urbakh}}, \ and\ \bibinfo {author} {\bibfnamefont {E.}~\bibnamefont
  {Meyer}},\ }\href {\doibase 10.1021/acsnano.5b05761} {\bibfield  {journal}
  {\bibinfo  {journal} {ACS Nano}\ }\textbf {\bibinfo {volume} {10}},\ \bibinfo
  {pages} {713} (\bibinfo {year} {2016})}\BibitemShut {NoStop}%
\bibitem [{\citenamefont {Kohn}\ and\ \citenamefont
  {Santoro}(2021{\natexlab{a}})}]{PRB_Kohn_2021}%
  \BibitemOpen
  \bibfield  {author} {\bibinfo {author} {\bibfnamefont {L.}~\bibnamefont
  {Kohn}}\ and\ \bibinfo {author} {\bibfnamefont {G.~E.}\ \bibnamefont
  {Santoro}},\ }\href {\doibase 10.1103/PhysRevB.104.014303} {\bibfield
  {journal} {\bibinfo  {journal} {Phys. Rev. B}\ }\textbf {\bibinfo {volume}
  {104}},\ \bibinfo {pages} {014303} (\bibinfo {year}
  {2021}{\natexlab{a}})}\BibitemShut {NoStop}%
\bibitem [{\citenamefont {Kohn}\ and\ \citenamefont
  {Santoro}(2021{\natexlab{b}})}]{arXiv_Kohn_2021}%
  \BibitemOpen
  \bibfield  {author} {\bibinfo {author} {\bibfnamefont {L.}~\bibnamefont
  {Kohn}}\ and\ \bibinfo {author} {\bibfnamefont {G.~E.}\ \bibnamefont
  {Santoro}},\ }\href@noop {} {\  (\bibinfo {year} {2021}{\natexlab{b}})},\
  \Eprint {http://arxiv.org/abs/2107.02807} {arXiv:2107.02807} \BibitemShut
  {NoStop}%
\bibitem [{\citenamefont {Anderson}(1961)}]{Anderson_PR61}%
  \BibitemOpen
  \bibfield  {author} {\bibinfo {author} {\bibfnamefont {P.~W.}\ \bibnamefont
  {Anderson}},\ }\href {\doibase 10.1103/PhysRev.124.41} {\bibfield  {journal}
  {\bibinfo  {journal} {Phys. Rev.}\ }\textbf {\bibinfo {volume} {124}},\
  \bibinfo {pages} {41} (\bibinfo {year} {1961})}\BibitemShut {NoStop}%
\bibitem [{\citenamefont {Hewson}(1997)}]{Hewson_kondo:book}%
  \BibitemOpen
  \bibfield  {author} {\bibinfo {author} {\bibfnamefont {A.~C.}\ \bibnamefont
  {Hewson}},\ }\href {\doibase 10.1017/CBO9780511470752} {\emph {\bibinfo
  {title} {The {K}ondo problem to heavy fermions}}}\ (\bibinfo  {publisher}
  {Cambridge University Press},\ \bibinfo {year} {1997})\BibitemShut {NoStop}%
\bibitem [{\citenamefont {Costi}(2000)}]{PRL_Costi_2000}%
  \BibitemOpen
  \bibfield  {author} {\bibinfo {author} {\bibfnamefont {T.~A.}\ \bibnamefont
  {Costi}},\ }\href {\doibase 10.1103/PhysRevLett.85.1504} {\bibfield
  {journal} {\bibinfo  {journal} {Phys. Rev. Lett.}\ }\textbf {\bibinfo
  {volume} {85}},\ \bibinfo {pages} {1504} (\bibinfo {year}
  {2000})}\BibitemShut {NoStop}%
\bibitem [{\citenamefont {Chin}\ \emph {et~al.}(2010)\citenamefont {Chin},
  \citenamefont {Rivas}, \citenamefont {Huelga},\ and\ \citenamefont
  {Plenio}}]{JMath_Chin_2010}%
  \BibitemOpen
  \bibfield  {author} {\bibinfo {author} {\bibfnamefont {A.~W.}\ \bibnamefont
  {Chin}}, \bibinfo {author} {\bibfnamefont {{\'A}.}~\bibnamefont {Rivas}},
  \bibinfo {author} {\bibfnamefont {S.~F.}\ \bibnamefont {Huelga}}, \ and\
  \bibinfo {author} {\bibfnamefont {M.~B.}\ \bibnamefont {Plenio}},\ }\href
  {\doibase 10.1063/1.3490188} {\bibfield  {journal} {\bibinfo  {journal}
  {Journal of Mathematical Physics}\ }\textbf {\bibinfo {volume} {51}},\
  \bibinfo {pages} {092109} (\bibinfo {year} {2010})}\BibitemShut {NoStop}%
\bibitem [{\citenamefont {Takahashi}\ and\ \citenamefont
  {Umezawa}(1975)}]{ColPhen_Umezawa_1975}%
  \BibitemOpen
  \bibfield  {author} {\bibinfo {author} {\bibfnamefont {Y.}~\bibnamefont
  {Takahashi}}\ and\ \bibinfo {author} {\bibfnamefont {H.}~\bibnamefont
  {Umezawa}},\ }\href {\doibase 10.1142/S0217979296000817} {\bibfield
  {journal} {\bibinfo  {journal} {Collective Phenomena}\ }\textbf {\bibinfo
  {volume} {2}},\ \bibinfo {pages} {55} (\bibinfo {year} {1975})}\BibitemShut
  {NoStop}%
\bibitem [{\citenamefont {de~Vega}\ and\ \citenamefont
  {Ba\~nuls}(2015)}]{PRA_Vega_2015}%
  \BibitemOpen
  \bibfield  {author} {\bibinfo {author} {\bibfnamefont {I.}~\bibnamefont
  {de~Vega}}\ and\ \bibinfo {author} {\bibfnamefont {M.-C.}\ \bibnamefont
  {Ba\~nuls}},\ }\href {\doibase 10.1103/PhysRevA.92.052116} {\bibfield
  {journal} {\bibinfo  {journal} {Phys. Rev. A}\ }\textbf {\bibinfo {volume}
  {92}},\ \bibinfo {pages} {052116} (\bibinfo {year} {2015})}\BibitemShut
  {NoStop}%
\bibitem [{\citenamefont {Schwarz}\ \emph {et~al.}(2018)\citenamefont
  {Schwarz}, \citenamefont {Weymann}, \citenamefont {von Delft},\ and\
  \citenamefont {Weichselbaum}}]{PRL_Weichselbaum_2018}%
  \BibitemOpen
  \bibfield  {author} {\bibinfo {author} {\bibfnamefont {F.}~\bibnamefont
  {Schwarz}}, \bibinfo {author} {\bibfnamefont {I.}~\bibnamefont {Weymann}},
  \bibinfo {author} {\bibfnamefont {J.}~\bibnamefont {von Delft}}, \ and\
  \bibinfo {author} {\bibfnamefont {A.}~\bibnamefont {Weichselbaum}},\ }\href
  {\doibase 10.1103/PhysRevLett.121.137702} {\bibfield  {journal} {\bibinfo
  {journal} {Phys. Rev. Lett.}\ }\textbf {\bibinfo {volume} {121}},\ \bibinfo
  {pages} {137702} (\bibinfo {year} {2018})}\BibitemShut {NoStop}%
\bibitem [{\citenamefont {N\"u\ss{}eler}\ \emph {et~al.}(2020)\citenamefont
  {N\"u\ss{}eler}, \citenamefont {Dhand}, \citenamefont {Huelga},\ and\
  \citenamefont {Plenio}}]{PRB_Plenio_2020}%
  \BibitemOpen
  \bibfield  {author} {\bibinfo {author} {\bibfnamefont {A.}~\bibnamefont
  {N\"u\ss{}eler}}, \bibinfo {author} {\bibfnamefont {I.}~\bibnamefont
  {Dhand}}, \bibinfo {author} {\bibfnamefont {S.~F.}\ \bibnamefont {Huelga}}, \
  and\ \bibinfo {author} {\bibfnamefont {M.~B.}\ \bibnamefont {Plenio}},\
  }\href {\doibase 10.1103/PhysRevB.101.155134} {\bibfield  {journal} {\bibinfo
   {journal} {Phys. Rev. B}\ }\textbf {\bibinfo {volume} {101}},\ \bibinfo
  {pages} {155134} (\bibinfo {year} {2020})}\BibitemShut {NoStop}%
\bibitem [{\citenamefont {Wolf}\ \emph {et~al.}(2014)\citenamefont {Wolf},
  \citenamefont {McCulloch},\ and\ \citenamefont
  {Schollw\"ock}}]{PRB_Wolf_2014}%
  \BibitemOpen
  \bibfield  {author} {\bibinfo {author} {\bibfnamefont {F.~A.}\ \bibnamefont
  {Wolf}}, \bibinfo {author} {\bibfnamefont {I.~P.}\ \bibnamefont {McCulloch}},
  \ and\ \bibinfo {author} {\bibfnamefont {U.}~\bibnamefont {Schollw\"ock}},\
  }\href {\doibase 10.1103/PhysRevB.90.235131} {\bibfield  {journal} {\bibinfo
  {journal} {Phys. Rev. B}\ }\textbf {\bibinfo {volume} {90}},\ \bibinfo
  {pages} {235131} (\bibinfo {year} {2014})}\BibitemShut {NoStop}%
\bibitem [{\citenamefont {Wilson}(1975)}]{Wilson_RMP75}%
  \BibitemOpen
  \bibfield  {author} {\bibinfo {author} {\bibfnamefont {K.~G.}\ \bibnamefont
  {Wilson}},\ }\href {\doibase 10.1103/RevModPhys.47.773} {\bibfield  {journal}
  {\bibinfo  {journal} {Rev. Mod. Phys.}\ }\textbf {\bibinfo {volume} {47}},\
  \bibinfo {pages} {773} (\bibinfo {year} {1975})}\BibitemShut {NoStop}%
\bibitem [{\citenamefont {Bulla}\ \emph {et~al.}(2008)\citenamefont {Bulla},
  \citenamefont {Costi},\ and\ \citenamefont {Pruschke}}]{RMP_Bulla_2008}%
  \BibitemOpen
  \bibfield  {author} {\bibinfo {author} {\bibfnamefont {R.}~\bibnamefont
  {Bulla}}, \bibinfo {author} {\bibfnamefont {T.~A.}\ \bibnamefont {Costi}}, \
  and\ \bibinfo {author} {\bibfnamefont {T.}~\bibnamefont {Pruschke}},\ }\href
  {\doibase 10.1103/RevModPhys.80.395} {\bibfield  {journal} {\bibinfo
  {journal} {Rev. Mod. Phys.}\ }\textbf {\bibinfo {volume} {80}},\ \bibinfo
  {pages} {395} (\bibinfo {year} {2008})}\BibitemShut {NoStop}%
\bibitem [{\citenamefont {Schollw{\"o}ck}(2005)}]{Schollwoeck_RMP05}%
  \BibitemOpen
  \bibfield  {author} {\bibinfo {author} {\bibfnamefont {U.}~\bibnamefont
  {Schollw{\"o}ck}},\ }\href {\doibase 10.1103/RevModPhys.77.259} {\bibfield
  {journal} {\bibinfo  {journal} {Rev. Mod. Phys.}\ }\textbf {\bibinfo {volume}
  {77}},\ \bibinfo {pages} {259} (\bibinfo {year} {2005})}\BibitemShut
  {NoStop}%
\bibitem [{\citenamefont {Schollw{\"o}ck}(2011)}]{AoP_Schollwoeck_2011}%
  \BibitemOpen
  \bibfield  {author} {\bibinfo {author} {\bibfnamefont {U.}~\bibnamefont
  {Schollw{\"o}ck}},\ }\href {\doibase 10.1016/j.aop.2010.09.012} {\bibfield
  {journal} {\bibinfo  {journal} {Annals of Physics}\ }\textbf {\bibinfo
  {volume} {326}},\ \bibinfo {pages} {96} (\bibinfo {year} {2011})}\BibitemShut
  {NoStop}%
\bibitem [{\citenamefont {White}(1992)}]{White_PRL92}%
  \BibitemOpen
  \bibfield  {author} {\bibinfo {author} {\bibfnamefont {S.~R.}\ \bibnamefont
  {White}},\ }\href {\doibase 10.1103/PhysRevLett.69.2863} {\bibfield
  {journal} {\bibinfo  {journal} {Phys. Rev. Lett.}\ }\textbf {\bibinfo
  {volume} {69}},\ \bibinfo {pages} {2863} (\bibinfo {year}
  {1992})}\BibitemShut {NoStop}%
\bibitem [{\citenamefont {Haegeman}\ \emph {et~al.}(2016)\citenamefont
  {Haegeman}, \citenamefont {Lubich}, \citenamefont {Oseledets}, \citenamefont
  {Vandereycken},\ and\ \citenamefont {Verstraete}}]{PRB_Haegeman_2016}%
  \BibitemOpen
  \bibfield  {author} {\bibinfo {author} {\bibfnamefont {J.}~\bibnamefont
  {Haegeman}}, \bibinfo {author} {\bibfnamefont {C.}~\bibnamefont {Lubich}},
  \bibinfo {author} {\bibfnamefont {I.}~\bibnamefont {Oseledets}}, \bibinfo
  {author} {\bibfnamefont {B.}~\bibnamefont {Vandereycken}}, \ and\ \bibinfo
  {author} {\bibfnamefont {F.}~\bibnamefont {Verstraete}},\ }\href {\doibase
  10.1103/PhysRevB.94.165116} {\bibfield  {journal} {\bibinfo  {journal} {Phys.
  Rev. B}\ }\textbf {\bibinfo {volume} {94}},\ \bibinfo {pages} {165116}
  (\bibinfo {year} {2016})}\BibitemShut {NoStop}%
\bibitem [{\citenamefont {Paeckel}\ \emph {et~al.}(2019)\citenamefont
  {Paeckel}, \citenamefont {K{\"o}hler}, \citenamefont {Swoboda}, \citenamefont
  {Manmana}, \citenamefont {Schollw\"ock},\ and\ \citenamefont
  {Hubig}}]{AoP_Paeckel_2019}%
  \BibitemOpen
  \bibfield  {author} {\bibinfo {author} {\bibfnamefont {S.}~\bibnamefont
  {Paeckel}}, \bibinfo {author} {\bibfnamefont {T.}~\bibnamefont {K{\"o}hler}},
  \bibinfo {author} {\bibfnamefont {A.}~\bibnamefont {Swoboda}}, \bibinfo
  {author} {\bibfnamefont {S.~R.}\ \bibnamefont {Manmana}}, \bibinfo {author}
  {\bibfnamefont {U.}~\bibnamefont {Schollw\"ock}}, \ and\ \bibinfo {author}
  {\bibfnamefont {C.}~\bibnamefont {Hubig}},\ }\href {\doibase
  10.1016/j.aop.2019.167998} {\bibfield  {journal} {\bibinfo  {journal} {Annals
  of Physics}\ }\textbf {\bibinfo {volume} {411}},\ \bibinfo {pages} {167998}
  (\bibinfo {year} {2019})}\BibitemShut {NoStop}%
\bibitem [{\citenamefont {Fugger}\ \emph {et~al.}(2018)\citenamefont {Fugger},
  \citenamefont {Dorda}, \citenamefont {Schwarz}, \citenamefont {von Delft},\
  and\ \citenamefont {Arrigoni}}]{NJP_vonDelft_2018}%
  \BibitemOpen
  \bibfield  {author} {\bibinfo {author} {\bibfnamefont {D.~M.}\ \bibnamefont
  {Fugger}}, \bibinfo {author} {\bibfnamefont {A.}~\bibnamefont {Dorda}},
  \bibinfo {author} {\bibfnamefont {F.}~\bibnamefont {Schwarz}}, \bibinfo
  {author} {\bibfnamefont {J.}~\bibnamefont {von Delft}}, \ and\ \bibinfo
  {author} {\bibfnamefont {E.}~\bibnamefont {Arrigoni}},\ }\href {\doibase
  10.1088/1367-2630/aa9fdc} {\bibfield  {journal} {\bibinfo  {journal} {New
  Journal of Physics}\ }\textbf {\bibinfo {volume} {20}},\ \bibinfo {pages}
  {013030} (\bibinfo {year} {2018})}\BibitemShut {NoStop}%
\bibitem [{\citenamefont {{\v{Z}}itko}\ and\ \citenamefont
  {Pruschke}(2010)}]{NJP_Zitko_2010}%
  \BibitemOpen
  \bibfield  {author} {\bibinfo {author} {\bibfnamefont {R.}~\bibnamefont
  {{\v{Z}}itko}}\ and\ \bibinfo {author} {\bibfnamefont {T.}~\bibnamefont
  {Pruschke}},\ }\href {\doibase 10.1088/1367-2630/12/6/063040} {\bibfield
  {journal} {\bibinfo  {journal} {New Journal of Physics}\ }\textbf {\bibinfo
  {volume} {12}},\ \bibinfo {pages} {063040} (\bibinfo {year}
  {2010})}\BibitemShut {NoStop}%
\bibitem [{\citenamefont {Fishman}\ \emph {et~al.}(2020)\citenamefont
  {Fishman}, \citenamefont {White},\ and\ \citenamefont
  {Stoudenmire}}]{itensor}%
  \BibitemOpen
  \bibfield  {author} {\bibinfo {author} {\bibfnamefont {M.}~\bibnamefont
  {Fishman}}, \bibinfo {author} {\bibfnamefont {S.~R.}\ \bibnamefont {White}},
  \ and\ \bibinfo {author} {\bibfnamefont {E.~M.}\ \bibnamefont
  {Stoudenmire}},\ }\href@noop {} {\  (\bibinfo {year} {2020})},\ \Eprint
  {http://arxiv.org/abs/2007.14822} {arXiv:2007.14822} \BibitemShut {NoStop}%
\end{thebibliography}%

\end{document}